\documentclass [12pt] {article}
\usepackage [cp1251] {inputenc}
\usepackage {amsmath, amsfonts}
\usepackage {graphicx}
\newcommand {\beq} {\begin {equation}}
\newcommand {\eeq} {\end {equation}}
\newcommand {\m} [1] {\marginpar {*1}}
\newcommand {\vT} {v _ {\scriptscriptstyle T}}
\newcommand {\di} {\displaystyle}

\newcommand {\p} {\partial}

\begin {document}

\begin {center}
{ \noindent \large Piecewise continuous partition function method
\\} { \noindent \large in the theory of wave perturbations of
 inhomogeneous gas}

{ Vereshchagin D.A., Leble S.B., Solovchuk M.A. \\
\small 236041, Kaliningrad, Al. Nevsky str. 14., \\
\small Kaliningrad State University, \\
\small Theoretical Physics Department}
\end {center}

\begin{abstract}
The problem of wave disturbance propagation in rarefied gas in
gravity field is explored. The system of hydrodynamic-type
equations for a stratified gas in gravity field is derived from
BGK equation by method of piecewise continuous partition function.
The obtained system of the equations generalizes the Navier-Stokes
at arbitrary density (Knudsen numbers). The verification of the
model is made for a limiting case of a homogeneous medium. Results
are in the good agreement with experiment and former theories at
arbitrary Knudsen numbers.
\end{abstract}

\section*{Introduction}
There is a significant number of problems of gas dynamics at which
 it is necessary to use the mathematical apparatus
beyond the limits of traditional hydrodynamics of Navier - Stokes.
The hydrodynamics is valid under the condition for the Knudsen
number $Kn = {\it l}/L \ll 1 $, where $ {\it l} $ is a mean free
path, and $L $ is a characteristic scale of inhomogeneity of a
problem under consideration. The first work, in which wave
perturbations of a gas were investigated from the point of view of
more general kinetic approach, perhaps, is the paper of Wang of
Chang and Uhlenbeck \cite{chang}. Most consistently these ideas
are formulated in the work of Foch and Ford \cite{ford}. Such
general theory could be based on some kinetic approach, i.e.
Boltzman equation.

Numerous researches on a sound propagation in a homogeneous gas at
arbitrary Knudsen numbers were made \cite{meyer}~-~\cite{cheng}.
The investigations have shown, that at arbitrary Knudsen numbers
the behaviour of a wave differs considerably from ones predicted
on a basis of hydrodynamical equations of Navier - Stokes. These
researches have revealed two essential features: first,
propagating perturbations keep wave properties at larger values of
$Kn $, than it could be assumed on the basis of a hydrodynamical
description. Secondly, at $Kn \ge 1 $ such concepts as a wave
vector and frequency of a wave become ill-determined.

The case, when Knudsen number $Kn $ is non-uniform in space or in
time is more difficult for investigation and hence need more
simplifications in  kinetic equations or their model analogues. A
constructions of such approaches for analytical solutions based on
kinetic equation Bhatnagar -- Gross -- Krook (BGK) of
Gross-Jackson \cite{gross} in a case of exponentially stratified
gas were considered at \cite{veresc3,veresc4}

In this paper we would  develop and generalize  the method of a
piecewise continuous partition functions \cite{veresc3,veresc4} to
take into account the complete set of nonlinearities (?). We
consider the example of wave perturbations theory for a gas
stratified in gravity field so that the Knudsen number
exponentially depends on the (vertical) coordinate.

\section{Piecewise continuous partition function method}
The kinetic equation with the model integral of collisions in BGK
form looks like:
\begin {equation}
\label {q1} \frac {\partial f} {\partial t} + \vec v\frac
{\partial f} {\partial \vec r}- g\frac {\partial f} {\partial v_z}
= \nu\left (f _ {\it l}-f\right) \,
\end {equation}
here $f $ is the distribution function of a gas, $t $ is time, $
\vec v $ is velocity of a particle of a gas, $ \vec r $ is
coordinate,
$$
f_{\it l}(\vec r, \vec v, t) )=\frac{n}{\pi^{3/2}\vT^3}
\exp\left(-\frac{(\vec v-\vec U)^2}{\vT^2}\right)
$$
is the local-equilibrium distribution function, $H=kT/mg $ is the
so-called  height of a homogeneous atmosphere - a parameter of the
gas stratification, $ \vT =\sqrt{2kT/m} $ is the average thermal
speed of movement of particles of gas, $ \nu =\nu_0\exp (-z/H) $
is the effective frequency of collisions between particles of gas
at height $z $. It is supposed, that density of gas $n $, its
average speed $ \vec U = (u_x, u_y, u_z) $ and temperature $T $
are functions of time and coordinates.

Following the idea of the method of piecewise continuous
distribution functions let's search for the solution $ f $ of the
equations~(\ref{q2}) as combinations of two locally equilibrium
distribution functions, each of which gives the contribution in
its own area of velocities space:
\begin{equation}
\label{q2} f(t, \vec r, \vec V)=\left\{
\begin{array}{rcl}
 f^+&=&\di n^+ \left(\frac{m}{2\pi k T^+}\right)^{3/2}
\exp\left(-\frac{m(\vec V-\vec U^+)^2}{2 k T^+}\right)\ , \qquad
v_z\ge 0 \vspace{1mm}\\
  f^-&=&\di n^- \left(\frac{m}{2\pi k T^-}\right)^{3/2}
\exp\left(-\frac{m(\vec V-\vec U^-)^2}{2 k T^-}\right)\ ,
\qquad v_z< 0\\
\end{array}
\right.
\end{equation}
here $n^{\pm}, U^{\pm}, T^{\pm}$ are parameters of locally
equilibrium distributions functions. Geometry of break, that is
the area, in which various functions operate, is determined by
geometry of a problem.

Thus, a set of the parameters determining a state of the perturbed
gas is increased twice. The increase of the number of parameters
of distribution function (\ref {q2}) results in that the
distribution function generally differs from a local-equilibrium
one and  describes deviations from hydrodynamical regime. In the
range of small Knudsen numbers $ {\it l}<< L $ we have $ n^+ =
n^-, U^+ = U^-, T^+ = T^-$ and distribution function (\ref{q2})
tends to local-equilibrium one, reproducing exactly the
hydrodynamics of Navier-Stokes. In the range of big Knudsen
numbers the formula~(\ref {q2}) gives solutions of collisionless
problems. Similar ideas have resulted successfully in a series of
problems. For example, in papers \cite{lees}~-~\cite{schidl} a
method of piecewise continuous partition function was demonstrated
for the description of flat and cylindrical (neutral and plasma)
Kuette flows ~\cite{lees}~-~\cite{schidl}. Thus for a flat problem
the surface of break in the velocity space was determined by a
natural condition $V_z=0 $, and in a cylindrical case $V_r=0 $,
where $V_z $ and $V_r $ are, accordingly, vertical and radial
components of velocity of particles. Similar problem was solved by
perturbations caused by pulse movement of plane~\cite{schidl,
kostom}. Solving a problem of of a shock wave
structure~\cite{schidl, mott, nambu} the solution was represented
as a combination of two locally equilibrium functions, one of
which determines the solution before front of a wave, and another
- after. In the problem  of  condensation/evaporation of drops of
a given size~\cite{sampson, ivchenko} a surface break was
determined by so-called "cone of influence", thus all particles
were divided into two types: flying "from a drop" and flying "not
from a drop".

The similar approach was developed for a description of a
nonlinear sound in stratified gas ~\cite{leble2, veresc4}.

The idea of a method of two-fold distribution functions given by
 ~(\ref {q2})  is realized as follows. Let's multiply
equation BGK~(\ref {q1}) on a set of linearly independent
functions. In the one-dimensional case $ \vec U = (0, 0, U_z) $
the following set is used:
 \begin{equation}
\label{q3}
\begin{array}{rclrrcl}
 \varphi_1&=&m\ ,
 & \varphi_4&=& m(V_z - U_z)^2\ ,\vspace{2mm} \\
 \varphi_2&=&m V_z\ ,
 & \varphi_5&=&\di \frac{1}{2} m(V_z-U_z)|\vec V - \vec U|^2\ ,\vspace{2mm}\\
 \varphi_3&=&\di \frac{1}{2}m |\vec V - \vec U|^2\ ,
 & \varphi_6&=&\di \frac{1}{2} m(V_z-U_z)^3\ .\\
\end{array}
\end{equation}

Let's define a scalar product:
\begin{equation}
\label{q4}
 <\varphi_n,f> \equiv <\varphi_n>
 \equiv \int d\vec v\: \varphi_n(t,z,\vec V) f(t,z,\vec V)\ .
\end{equation}
\begin{equation}
\label{q5}
\begin{array}{ll}
 <\varphi_1> = m<1>=\rho(t, z) \ ,
 & <\varphi_4> = m<(V_z - U_z)^2> = P_{zz}\ ,\vspace{1mm} \\
 <\varphi_2>= m <V_z> = \rho U_z\ ,
 & <\varphi_5> =\di \frac{1}{2} m <(V_z-U_z)|\vec V - \vec U|^2> = q_z\
,\vspace{1mm}\\
 <\varphi_3> = \di \frac{1}{2}m <|\vec V - \vec U|^2> = e\ ,
 & <\varphi_6>= \di \frac{1}{2} m <(V_z-U_z)^3> = \bar q_z\ .\\
\end{array}
\end{equation}
Here $\rho$ is density, $\rho U_z$ is specific momentum, $e$ is
internal energy per unit mass of the gas, $P_{zz} = P + \pi_{zz}$
is the diagonal component of the strain tensor ($P$ is pressure,
$\pi_{zz}$ is component of strain tensor), $q_z$ is a vertical
component of a heat flow, $\bar q_z$ is the new parameter having
meaning of a heat flow.

Multiplying equation (\ref {q1}) on eigen functions~(\ref {q3}) we
  obtain the system of differential equations:
\begin{equation}
\label{q13}
\begin{array}{l}
 \di
 \frac{\p}{\p t}\rho + \frac{\p}{\p z}(\rho U_z) = 0
\ ,\vspace{2mm} \\
 \di
 \frac{\p}{\p t}U_z + U_z \frac{\p}{\p z}U_z +
 \frac{1}{\rho} \frac{\p}{\p z}(P + \Pi_{zz}) + g = 0
\ ,\vspace{2mm} \\
 \di
 \frac{\p}{\p t}e + U_z \frac{\p}{\p z}e +
 (e + P + \Pi_{zz})\frac{\p}{\p z}U_z + \frac{\p}{\p z}q_z = 0
\ ,\vspace{2mm} \\
 \di
 \frac{\p}{\p t}(P + \Pi_{zz}) + U_z \frac{\p}{\p z}(P + \Pi_{zz}) +
 3 (P + \Pi_{zz})\frac{\p}{\p z}U_z + 2 \frac{\p}{\p z}\bar q_z =
 - \nu (\Pi_{zz} + P - \rho \theta)
\ ,\vspace{2mm} \\
  \di
  \frac{\p}{\p t}q_z + U_z \frac{\p}{\p z}q_z +
 2 (q_z + \bar q_z)\frac{\p}{\p z}U_z - \frac{1}{\rho} (e + P + \Pi_{zz})
\frac{\p}{\p z}(P + \Pi_{zz}) +
 \frac{\p}{\p z}J_1 = - \nu q_z
\ ,\vspace{2mm} \\
 \di
 \frac{\p}{\p t}\bar q_z + U_z \frac{\p}{\p z}\bar q_z +
 4 \bar q_z \frac{\p}{\p z}U_z -
 \frac{3}{2\rho} (P + \Pi_{zz}) \frac{\p}{\p z}(P + \Pi_{zz}) +
 \frac{\p}{\p z}J_2 = - \nu \bar q_z\ ,
\end{array}
\end{equation}

\begin{equation}
\label{q14}
\begin{array}{ll}
 \di \text{where}\quad  J_1 = <(V_z - U_z)^2 (\vec V - \vec  U)^2> \ ,
 \quad &
 J_2 = <(V_z - U_z)^4>\ . \\
\end{array}
\end{equation}
The obtained system (\ref{q13}) of the equations according to the
derivation scheme is valid at all frequencies of collisions and
within the limits of high frequencies should transform to the
hydrodynamic equations. It is a system of hydrodynamical type and
generalizes the classical equations of a viscous fluid on any
density, down to a free-molecule flow. However, the system (\ref
{q13}) is not closed yet. It is necessary to add equations of
state $ P=P (\rho, T) $ and $e=e (\rho, T) $. Except for that it
is necessary to present values of two integrals $J_1 $ and $J_2 $
as functions of thermodynamic parameters of the system  (\ref
{q13}).

Let's evaluate integrals (\ref {q14}) directly, plugging the
function (\ref {q2}). We estimate the functions  $ \di \frac {U ^
{\pm}} {V_T ^ {\pm}} \ $ as small, that corresponds to small Mach
numbers $M = max |\vec v|/ v_T $. Values of integrals $J_1 $ and
$J_2 $ within the specified approximation looks as
\begin{equation}
\label{q16}
\begin{array}{c}
 \di J_1 = \frac{5}{16}  (n^+ {V_T^+}^4 + n^- {V_T^-}^4) +
 \frac{3}{2 \sqrt{\pi}}  [n^+ {V_T^+}^3 (U^+ - U) - n^- {V_T^-}^3 (U^- -
U) ]
\ ,\vspace{2mm} \\
 \di J_2 = \frac{3}{16}  (n^+ {V_T^+}^4 + n^- {V_T^-}^4 ) +
 \frac{1}{\sqrt{\pi}}  [n^+ {V_T^+}^3 (U^+ - U) - n^- {V_T^-}^3 (U^- -
U) .
\end{array}
\end{equation}
Let's express parameters of the two-fold distribution function
(\ref {q2}) through the thermodynamic ones and substitute the
result into the expression (\ref {q16}). To solve the specified
problem we shall use a method of perturbations with the small
parameter $ \frac {max U ^ {\pm}} {V_T ^ {\pm}} $, expressing
\begin{equation} \label{UV}
\begin{array}{c}
   U^+ = - \frac{\sqrt{3 \pi}}{4} \sqrt{\frac{\rho}{e}}
 (P_{zz} - \frac{2}{3} e ) +
 \frac{1}{5 e} (5 U e + 3 q) , \\
  \rho^+ = \rho + \frac{3 \rho}{4 e} (P_{zz} - \frac{2}{3} e ) +
 \frac{3}{20} \sqrt{3 \pi}  (\frac{\rho}{e})^{3/2} q \\
  U^- = \frac{\sqrt{3 \pi}}{4} \sqrt{\frac{\rho}{e}} (P_{zz}
- \frac{2}{3} e ) +
 \frac{1}{5 e} (5 U e + 3 q), \\
   \rho^- = \rho + \frac{3 \rho}{4 e} (P_{zz} - \frac{2}{3} e ) -
 \frac{3}{20} \sqrt{3 \pi}  (\frac{\rho}{e} )^{3/2} q, \\
  V_T^+ = 2 \sqrt{\frac{e}{3 \rho}} -
 \frac{1}{12} \sqrt{\frac{3}{\rho e}}  (P_{zz} - \frac{2}{3} e ) -
 \frac{3}{10} \frac{q}{e} \sqrt{\pi} ,\\
 V_T^- = 2 \sqrt{\frac{e}{3 \rho}} +
 \frac{1}{12} \sqrt{\frac{3}{\rho e}} (P_{zz} - \frac{2}{3} e ) +
 \frac{3}{10} \frac{q}{e} \sqrt{\pi}.
\end{array}
\end{equation}
Plugging the values of (\ref{UV}) into (\ref {q16}) one obtain the
values of $J_{1,2}$ in the first order :
\begin{equation}
\label{q21}
 \di
 J_1 = \frac{10}{9} \frac{e^2}{\rho} +
 \frac{61}{18} \frac{e}{\rho} \left(P_{zz} - \frac{2}{3} e\right) \ , \qquad
 J_2 = \frac{2}{3} \frac{e^2}{\rho} +
 \frac{13}{6} \frac{e}{\rho} \left(P_{zz} - \frac{2}{3} e\right) \ .
\end{equation}

\section{Limiting case of gas oscillations at high frequencies of
collisions (small Knudsen numbers).}

Let us consider a system in the hydrodynamical limit ($\nu \to
\infty$). It follows from the last three equations of the system
(\ref{q13}) that  the orders of values relate as $max\{\Pi_{zz},
q_z, \bar q_z\} \sim \nu^{-1} \max\{\rho, U_z, e, P\}$. Next
assume $\nu^{-1}=0$ in the zero order by the parameter $\nu^{-1}$.
One hence have $\Pi_{zz}=0, q_z=0, \bar q_z=0$ at the l.h.s. and
at the r.h.s. of forth equation of the system $P=\rho \theta = 2
e/3$. Substituting mentioned limits in the first three equations
of the system (\ref{q13}) we obtain a system of Euler equations of
a liquid in gravity field:
\begin{equation}
\label{q22}
\begin{array}{l}
 \di \frac{\p}{\p t}\rho + \frac{\p}{\p z}(\rho U_z) = 0
\vspace{2mm} \\
 \di \frac{\p}{\p t}U_z + U_z \frac{\p}{\p z}U_z +
 \frac{2}{3 \rho} \frac{\p}{\p z}e + g = 0
\vspace{2mm} \\
 \di \frac{\p}{\p t}e + U_z \frac{\p}{\p z}e +
 \frac{5}{3} e \frac{\p}{\p z}U_z = 0\ .
\end{array}
\end{equation}
The functions $\{\Pi_{zz}, q_z, \bar q_z\} \sim \nu^{-1} \{\rho,
U_z, e, P\}$ belong to  the next order of the parameter
$\nu^{-1}$. Then from the last three equations of the system,
(\ref{q13}) taking into account the
 equation of state $P=\rho \theta$, one obtains following
relations
\begin{equation}
\label{q23}
 \di
 \pi_{zz} = - \frac{8}{9 \nu(z)} e \frac{\p}{\p z}U_z \ , \qquad
 q_z = - \frac{10}{9 \nu(z)}
 \frac{\p}{\p z}\left(\frac{e}{\rho}\right) \ , \qquad
 \bar q_z = - \frac{2}{3 \nu(z)}
 \frac{\p}{\p z}\left(\frac{e}{\rho}\right) \ .
\end{equation}

Further substituting (\ref{q23}) in the first three equations of
the system (\ref{q13}) we obtain
\begin{equation}
\label{q24}
\begin{array}{l}
 \di \frac{\p}{\p t}\rho + \frac{\p}{\p z}(\rho U_z) = 0
\vspace{2mm} \\
 \di \frac{\p}{\p t}U_z + U_z \frac{\p}{\p z}U_z +
 \frac{2}{3 \rho} \frac{\p}{\p z}e + g -
 \frac{8}{9 \rho} \frac{\p}{\p z}\left(\frac{e}{\nu} \frac{\p}{\p z}U_z\right) =
0
\vspace{2mm} \\
 \di \frac{\p}{\p t}e + U_z \frac{\p}{\p z}e +
 \frac{5}{3} e \frac{\p}{\p z}U_z -
 \frac{10}{9} \frac{\p}{\p z} \left(\frac{e}{\nu} \frac{\p}{\p z}
\frac{e}{\rho}\right) -
 \frac{8}{9} \frac{e}{\nu} \left(\frac{\p}{\p z}U_z\right)^2 = 0\ .
\end{array}
\end{equation}
System (\ref{q24}) is the system of equation of a non-ideal
liquid, to compare it with the  Navier - Stokes equations, we
continue the evaluation of viscosity factor and coefficient of
heat conductivity. Expressions for strain tensor and heat flow
tensor in one-dimensional hydrodynamics take a form
$$
 \di
 \pi_{zz} = - \frac{4}{3} \eta \frac{\p}{\p z}U_z \ , \qquad
 q_z = - \frac{2}{3} \kappa \frac{\p}{\p z} \theta \ ,
$$
where $\eta$ is the viscosity factor, and $\kappa$ is the
coefficient of heat conductivity. Comparing mentioned expressions
with corresponding items in equations (\ref{q24}) we obtain
\begin{equation}
\label{q25}
 \di
 \eta = - \frac{n_0 k T_0}{\nu} \ , \qquad
 \kappa = - \frac{5}{2} \frac{n_0 k T_0}{\nu} \ ,
\end{equation}
that coincides with the well known relations, given, for example,
in \cite{kerzon}. Finding the Prandtl number, taking into account
of molecular thermal capacity of the ideal gas under constant
pressure $C_p=5/2$, we obtain
$$
 \text{Pr} = \frac{\eta C_p}{\kappa} = 1 \ ,
$$
that do not coincide with the Prandtl number of ideal gas
($\text{Pr}_{id}=2/3$). The wrong Prandtl number is the main
disadvantage of BGK model, that, however, can be removed by
changing to the more exact models of collision integral, for
example, of Gross-Jackson \cite{gross}.

\section{Linearized system of the equations. Dispersion relation.}

For a closure of the system we use the  equation of state of ideal
gas. Linearized system of the equations (\ref{q22}) is given by
\begin{equation}
\label{q27}
\begin{array}{l}
 \di
 \frac{\p}{\p t}\rho + V_T \frac{\p}{\p z}U_z = 0 \ ,\vspace{2mm} \\
 \di
 \frac{\p}{\p t}U_z + \frac{1}{2} V_T \frac{\p}{\p z}(\rho + T + \Pi_{zz})
 = 0 \ ,\vspace{2mm} \\
 \di
 \frac{\p}{\p t}T + \frac{1}{3} V_T \frac{\p}{\p z}(2 U_z +  3 q_z)  = 0 \
,\vspace{2mm} \\
 \di
 \frac{\p}{\p t} \Pi_{zz} + \frac{1}{3} V_T \frac{\p}{\p z}(4 U_z -3 q_z + 9
\bar  q_z)
= - \nu \Pi_{zz} \ ,\vspace{2mm} \\
 \di
 \frac{\p}{\p t}q_z + \frac{1}{36} V_T \frac{\p}{\p z}(30 T + 31 \Pi_{zz})
 = - \nu q_z \ ,\vspace{2mm} \\
 \di
 \frac{\p}{\p t}\bar q_z + \frac{1}{12} V_T \frac{\p}{\p z}(6 T + 7 \Pi_{zz})
 = - \nu \bar q_z\ .
\end{array}
\end{equation}
For convenience we would introduce new notations $n_i $ for
hydrodynamical variables: $n_1 = \rho $, $n_2 = U_z $, $ n_3=T $,
$n_4 =\Pi _ {zz} $, $n_5=q_z $, $n_6 =\bar q_z $. The solution of
system (\ref {q27}) we search as
\begin{equation}
\label{q28} n_i=a_i exp(-iwt+ i k_z z),
\end{equation}
where $w $ is frequency of a wave, $k_z $ - the vertical component
of a wave vector.

Substituting (\ref {q28}) in (\ref {q27}),one obtains  a system of
the homogeneous algebraic equations with constant coefficients
which solution exists if
\begin{equation}
\label{q31} \frac {18}{125}  \tilde{k}^6 + \left( \frac 35 r^2 -
\frac {39}{25} - \frac {48}{25} i r \right) \tilde{k}^4 +
 \left( -i r^3 - \frac {24}{5} r^{2} + \frac {
23}{3} i r + \frac {58}{15} \right) \tilde{k}^{2} + i r^3 - 1 -3i
r + 3 r^{2}=0
\end{equation}
Here the dimensionless wave number $ \widetilde k=k C_0/w $ and
the Reynolds number $r =\nu/w $ are introduced, where $ C_0=
\sqrt{\frac 56} V_T$ - sound speed in Euler's approximation. The
Reynold's number $r $ and the Knudsen number are obviously linked:
$$ Kn =\frac {\lambda} {\lambda_b} = \frac {w} {\nu} \frac {V_T} {2 \pi C_0} =
\sqrt {\frac 65} \frac {1} {2 \pi r}$$.

Let $ \widetilde {k} = \beta + i\alpha, $ then
$$n_i=a_i exp(-iw(t - \frac {\beta}{C_0} z)) exp(-w \frac {\alpha}{C_0} z)$$
 and the real part $
\beta=C_0/C,$ $ \alpha $ - the factor of attenuation.

\section{The joint account of three modes.}
The basic Fourier component solution of the system (\ref {q27}) we
shall search as a superposition of three plane waves
\begin{equation}
\label{trm1}
 \di n_i=A_i^1 exp(-iwt+ i k_1 z) + A_i^2 exp(-iwt+ i k_2
z) + A_i^3 exp(-iwt+ i k_3 z), \
\end{equation}
where $k_j, \quad j=1,2,3$, are solutions of the dispersion
equation (\ref{q31}) correspondent to the modes.

Substituting (\ref{trm1}) into the linearized system, we express
$A_2^j,A_3^j,A_4^j,A_5^j,A_6^j$ through $A_1^j \equiv A^j $. For
$A_2^j,A_3^j$ we have:
\begin{equation}
\label{trm2}
\begin{array}{l}
 \di A_2^j =\frac {wA^j }{k_j V_T}  \ , \\
 \di A_3^j = \frac {A^j  \left( -31\,{V_T}^{2} k_j^2 + 24 iw^2 - 24 \nu
w + 62w^2 \right) }{ V_T^2 k_j^2 + 36 iw^2 - 36 \nu w }  \ . \\
\end{array}
\end{equation}
To determine the coefficients $A^1, A^2, A^3$ we should choose
boundary conditions. We consider a problem  in half-space and the
reflection of molecules from a plane as a diffuse  one \cite
{back2}. The boundary condition for the distribution function
looks as
$$
\di f(z=0,\vec V,t)= \frac {n}{\pi^{3/2}V_T^3} exp\{-\frac{(\vec V-\vec U_0
e^{-iwt})^2}{V_T^2}\} \quad \text{by} \  V_z>0 . \
$$
Here $U_0 $ stands for an amplitude of the hydrodynamic velocity
oscillations. For $\frac {U_0}{V_T} \ll 1 $ we have:
$$ \di  \varphi (z=0,\vec V,t)= \frac {f-f^{(0)}}{f^{(0)}}  \sim 2
\frac {U_0}{V_T} e^{-iwt} \quad \text{by} \  V_z>0 . \ $$

For hydrodynamical variables on the bound we obtain
\begin{equation}
\label{trm3}
\begin{array}{l}
\di \rho(z=0,t) = < \varphi(z=0,\vec V,t) > = \frac
{1}{\pi^{3/2}V_T^3} \int d \vec V \varphi(z=0,\vec V,t)
e^{-V^2/V_T^2} = \frac {U_0}{\sqrt \pi} e^{-iwt}
 , \ \\
\di U_z(z=0,t)=< \frac {V_Z}{V_T} \varphi(z=0,\vec V,t) > = \frac
{U_0}{2} e^{-iwt}
 , \ \\
\di  T(z=0,t)= < \frac {V^2}{V_T^2} \varphi(z=0,\vec V,t) > =
\frac {U_0}{\sqrt \pi} e^{-iwt}.\
\end{array}
\end{equation}

Substituting the values of (\ref{trm2}) into (\ref{trm1}) and
comparing right-hand sides of expression (\ref{trm1}) and
(\ref{trm3}) we obtain the system of equations in variables $A^j$.
Solving given system of equations we obtain variables
$A^1,A^2,A^3$.

In experiment acoustic pressure perturbation amplitude is
measured. Appropriate combination of the basic variables for the
pressure it is given by the formula
\begin{equation}
\label{trm6} \di P(z,t) e^{iwt}= \rho' + T' = (A^1+A_3^1) e^{ik_1
z} + (A^2+A_3^2) e^{ik_2 z} + (A^3+A_3^3) e^{ik_3 z} \
\end{equation}
The real part of this expression relates to experiment. In
 fig.1a) the real part of this expression is represented at
r=0.2, where $ \widetilde {Z}= z w/C_0 $ - dimensionless
coordinate. The attenuation factor $\alpha$ is determined as a
slope ratio of the diagram of the logarithm of amplitude of
pressure depending upon distance between oscillator and the
receiver. It is illustrated in fig. 1 b).

\includegraphics[height=7.2cm]{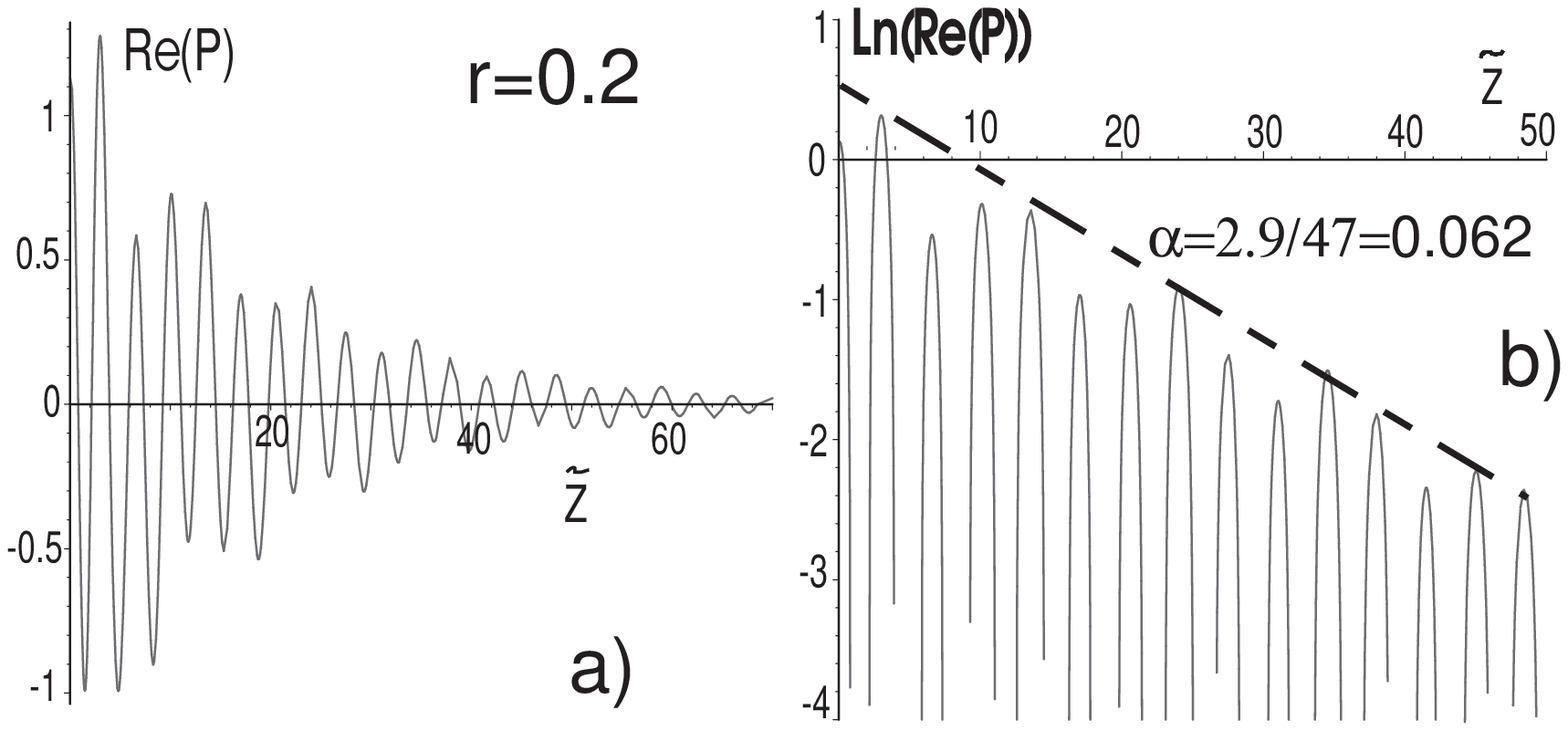}
\begin{center}
Fig. 1. Finding of attenuation factor
\end{center}

\section{Comparison with with experimant and results
of other evaluations.} In figures 2,3 a comparison of theoretical
results of the sound propagation parameters with experimental data
\cite{meyer,green} is made.

 The dispersion relation~(\ref {q31}) represents the
binary cubic equation with variable coefficients. The exact
analytical solution by the formula Cardano is very huge and
therefore we do not show it in this paper. At $ r \to 0 $ (free
molecule flow) we start from the propagation velocity by the
formula
$$ C_0/C=0.54+0.15r^2+0(r^4)$$.

\includegraphics[height=8cm]{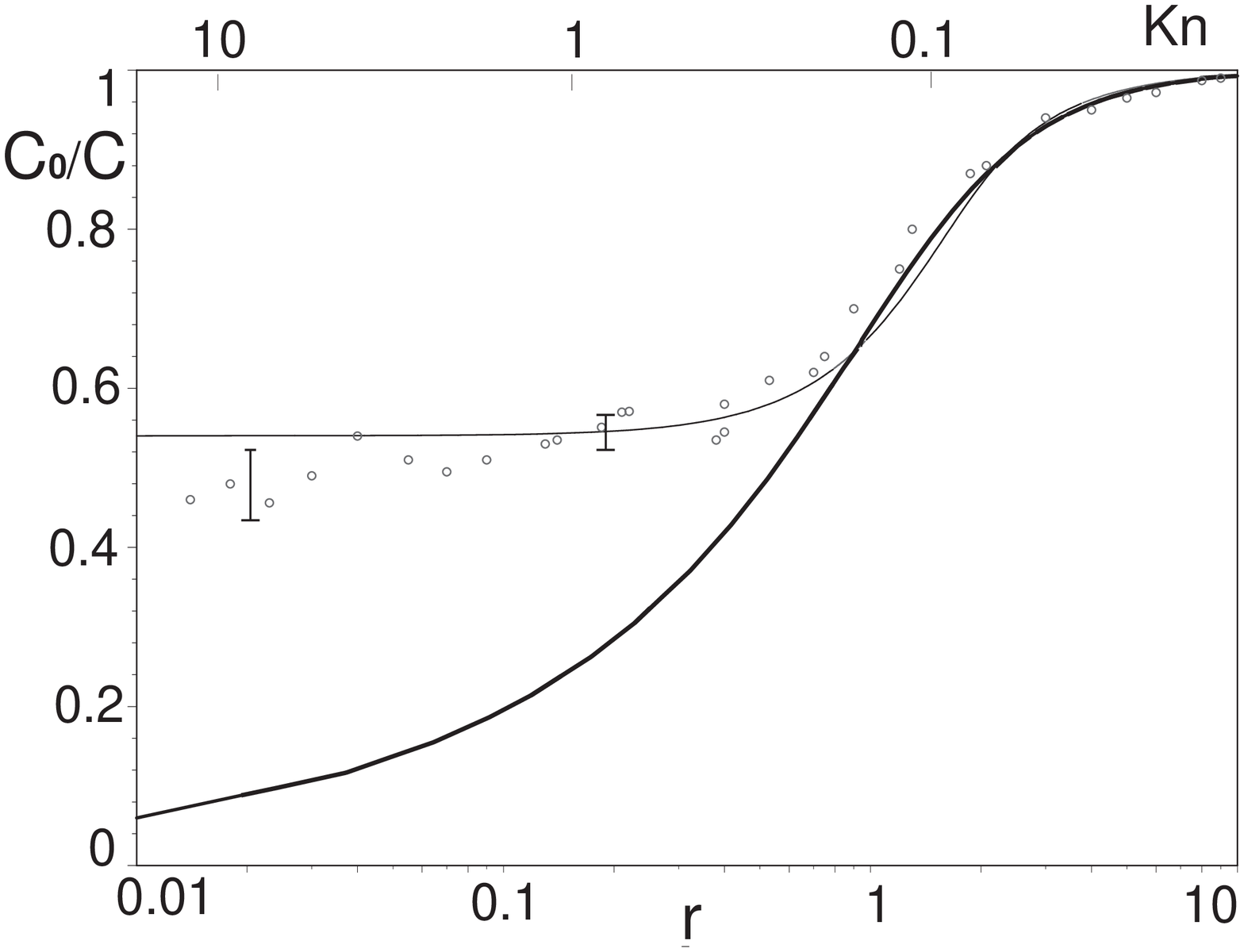}
\begin{center}
Fig.2. The propagation velocity. Thick line - Navier-Stokes, thin
line - this paper, circles - measurements in Argon (Greenspan,
Meyer-Sessler).
\end{center}

The attenuation factor is determined graphically as shown in the
fig. 1. Therefore we cannot introduce analytical expression.

\includegraphics[height=8cm]{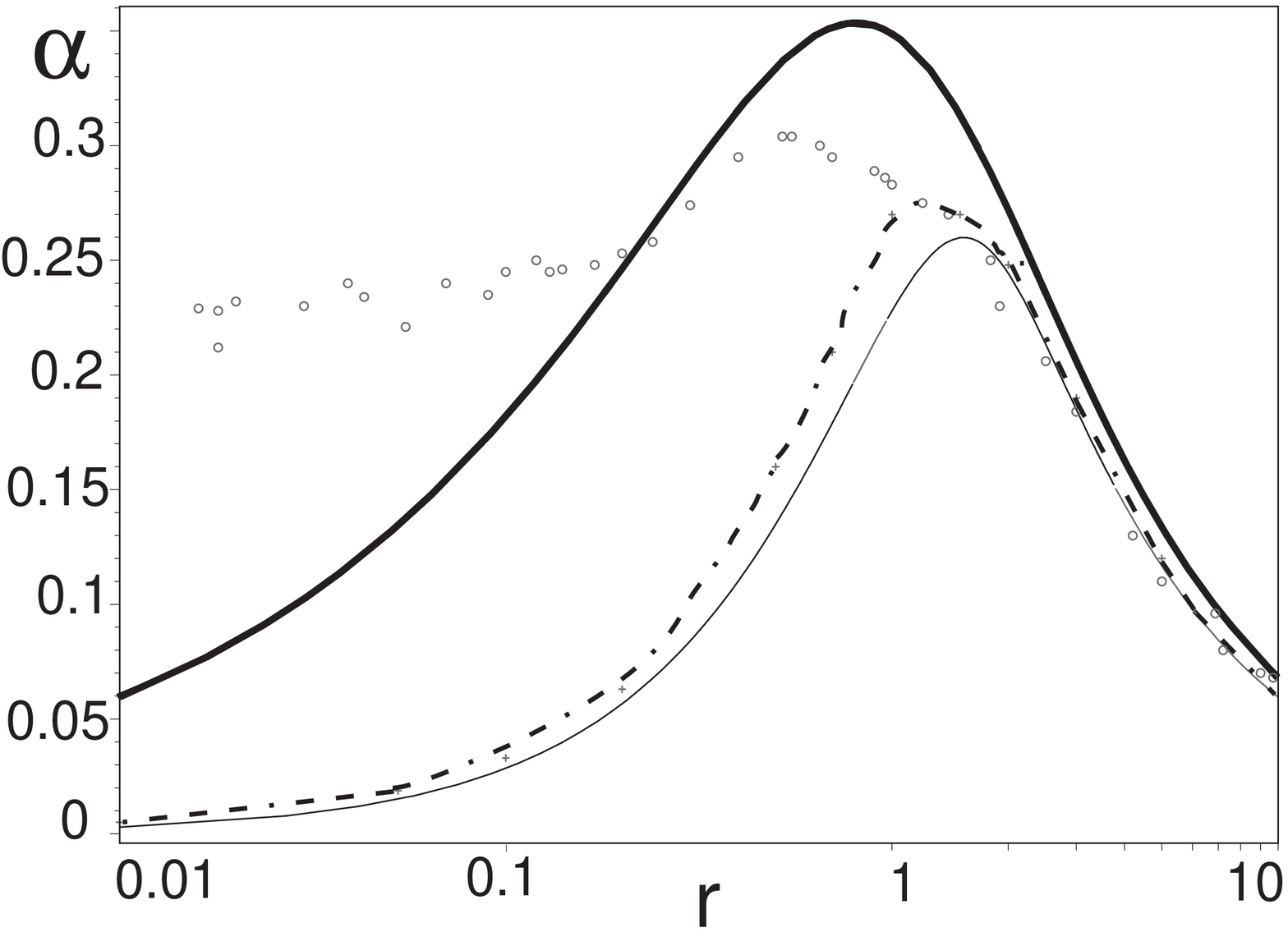}
\begin{center}
Fig.3. Attenuation in Argon. thick line - Navier-Stokes. thin line
- this paper(sound-wave). dotted line - the joint account of three
modes
\end{center}
Results for phase speed give the good consistency  with the
experiments. As we see, the account of three modes allows us to
enter  further the area of intermediate Knudsen numbers.

In figures 4, 5 a comparison of our results of numerical
calculation of dimensionless sound speed and attenuation factor
depending on $r$ is carried out with the results of the other
authors.

\begin{center}
\includegraphics[height=8cm]{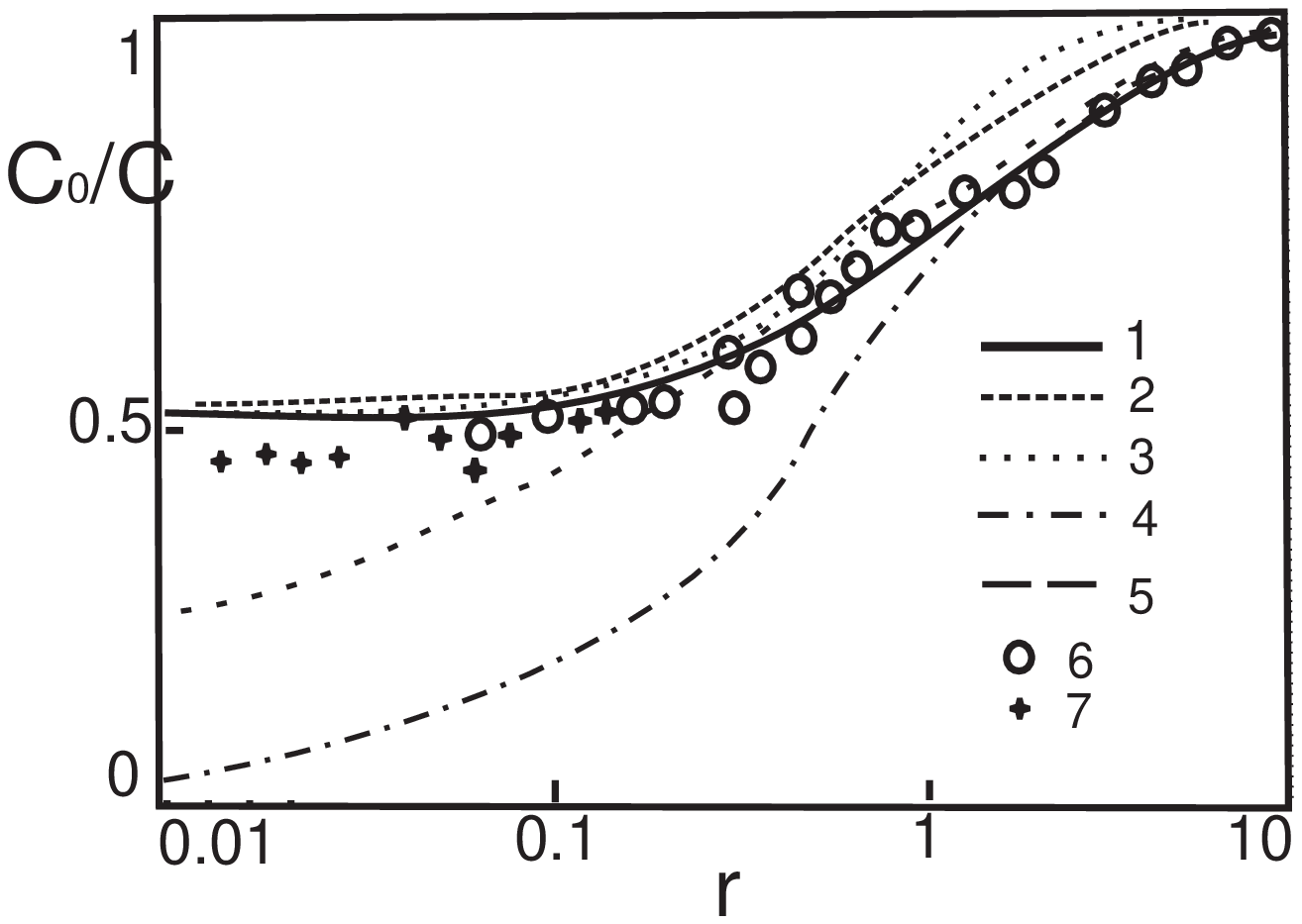}
\end{center}
\begin{center}
Fig. 4. A comparison of BGK model results with experimental data.
The propagation velocity. 1 - Present work, 2 - Loyalka and Cheng
\cite{loyalka},3- Buckner and Ferziger \cite{back2},4-
Navier-Stokes theory, 5- Sirovich and Thurber \cite{sirov3},
6-experimental data of Greenspan, 7- experimental data of Meyer
and Sessler
\end{center}

\begin{center}
\includegraphics[height=9cm,width=10cm]{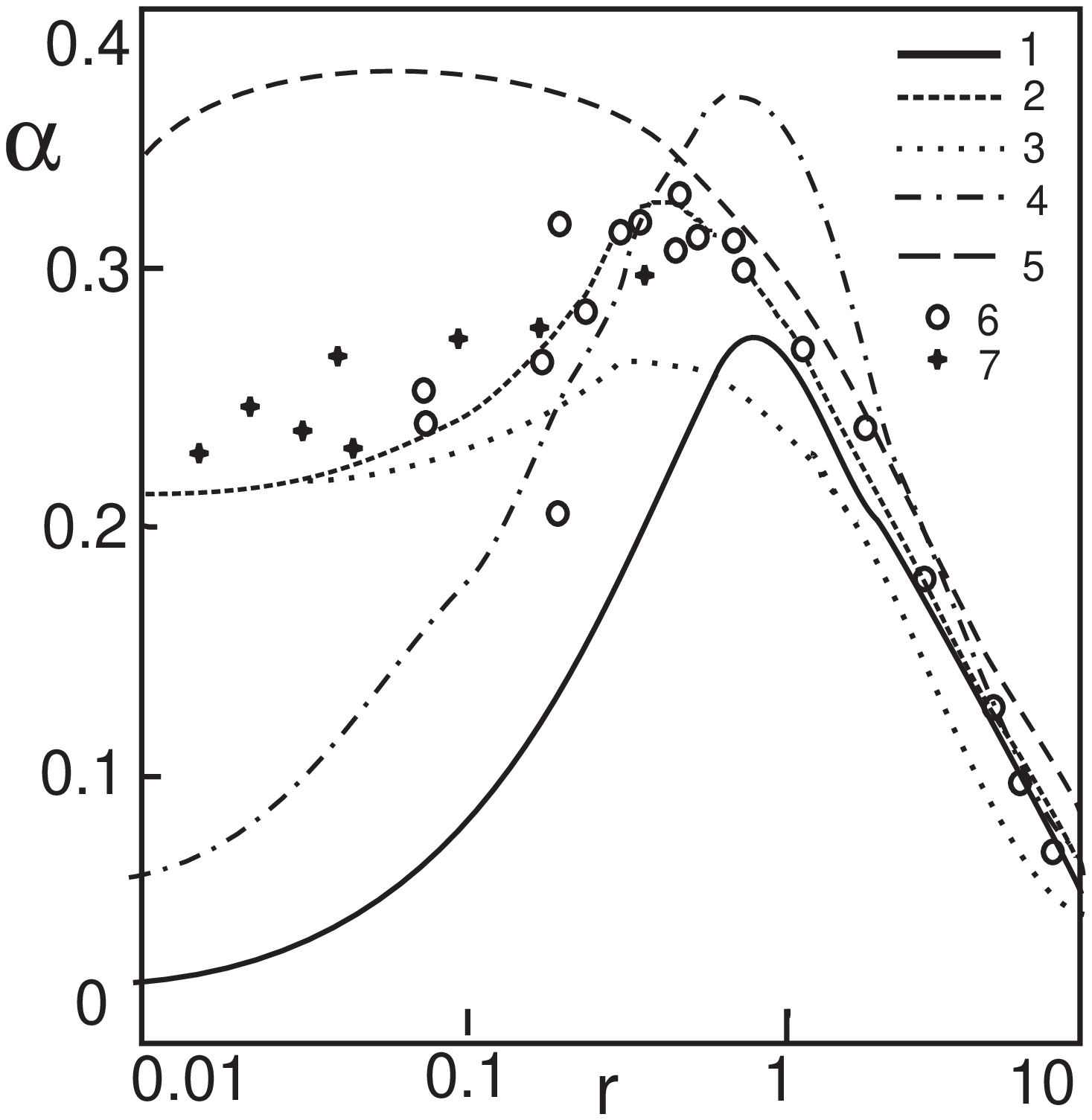}
\end{center}
\begin{center}
Fig. 5. A comparison of BGK model results with experimental data.
The attenuation factor. 1 - Present work, 2 - Loyalka and Cheng
\cite{loyalka}, 3- Buckner and Ferziger \cite{back2}, 4-
Navier-Stokes theory, 5- Sirovich and Thurber \cite{sirov3},
6-experimental data of Greenspan, 7- experimental data of Meyer
and Sessler
\end{center}

Lacks of the BGK model used in this article is that it gives
correct value of viscosity factor, but wrong value of coefficient
of heat conductivity. To the superior models of the Gross -
Jackson \cite{gross} this lack can be eliminated by transition.

At the solution of Boltzman equation the method of the Gross -
Jackson revealed sudden disappearance of discrete modes at some
values $r_c $ (\cite {back1}, \cite {back2}, \cite {sirov3}), and
with increase of number of the moments $r_c $ decreased.

For example, Buckner and Ferziger in the paper \cite{back2} have
shown, that for $r>1$ the solution is determined mainly by the
discrete sound mode and the dispersion relation may be used in
calculating the sound parameters. For $r < 1 $, the continuous
modes are important. The solution remains "wavelike", but it is no
longer a classical plane wave. In fact, the sound parameters are
depend on the position of the receiver.

Below $r_c $ the solution is represented as superposition of a
continuous spectrum of eigen functions, therefore the classical
understanding of a sound should be changed. The concept of a
dispersion relation is not applicable more.

\section{Conclusion}
The attenuation of sound at big Knudsen numbers is not
"damping"(due to intermolecular collisions), but rather "phase
mixing"(due to molekules which left the oscillator at different
phases arriving at the receiver at the same time)

The attenuation factor at big Knudsen numbers $ {Kn} > 1 $ is
modelled by the account of effects of a relaxation in integral of
collisions. The model of the Gross - Jackson at given N limits an
opportunity of the account external times of a relaxation (fast
attenuation) as essentially bases on a condition:
$$\lambda_i =\lambda _ {N+1}, \quad i > N+1 $$

Supreme times of a relaxation are assumed identical. It means,
that the inclusion of the supreme eigen functions $ \di \chi_i, \
\quad i \geq {N+1} $ is necessary, that would allow to move in the
range of higher Knudsen numbers.

In piecewise continuous partition function method the number of
waves is twice more, but restrictions on attenuation factor
remain.


\begin{thebibliography}{99}
\bibitem{chang}                  
Wang Chang C.S., Uhlenbeck G.E. Eng.Res.Ins., Univ. of
Michigan.Project M 999. Ann.Arbor., Michigan. (1952).
\bibitem{ford}
Foch D., Ford Jr.G.M. In ''Stadies in Statistical Mechanics'' (ed,
J. de Boer and G.E. Uhlenbeck), N.Holland,5. (1970). P.103-231.      
\bibitem{gross}                 %
Gross E.P., Jackson E.A. Phys. Fluids 1959 V.2 N4, P.432-441.
\bibitem{meyer}                  
Meyer E., Sessler G. Z.Physik. 149. (1957). P.15-39.
\bibitem{green}                  
Greenspan M. J.Acoust.Soc.Am., 28. $N^{\underline o} \ 4$. (1956)
P.644-648.
\bibitem{back1}                  
Buckner J.K., Ferziger J.H. Phys.Fluids.9. $N^{\underline o} \
12$. (1966). P.2309-2314.
\bibitem{back2}                  
Buckner J.K., Ferziger J.H. Phys.Fluids.9. $N^{\underline o} \
12$. (1966). P.2315-2322.
\bibitem{sirov2}                 
Sirovich L., Thurber J.K. Adv.Appl.Mech.,Supp.2. 1. (1963).
P.152-180.
\bibitem{sirov3}                 
Sirovich L., Thurber J.K. Acoust.Soc.Am.37. $N^{\underline o} \
2$. (1965). P.329-339.
\bibitem{sirov4}                 
Sirovich L., Thurber J.K.  J.Math.Phys.8. $N^{\underline o} \ 4$.
(1967). P.888-895.
\bibitem{loyalka}                
Loyalka S.K., Cheng T.S. Sound wave propagation in a rarefied gas.
Phys.Fluids.,22. $N^{\underline o} \ 5$. (1979). P.830-836.
\bibitem{cheng}                  
Cheng T.S., Loyalka S.K. Sound wave propagation in a rarefied gas.
II. Gross-Jackson model. Progress in Nuclear Energy. 8. (1981).
P.263-267.
\bibitem{aleks1}
Alekseev B. V. Physics-Uspekhi, Vol. 43(2000), N 6, P. 601-629.
\bibitem{aleks2}
Alekseev B. V. Physica A 216 459 (1995)
\bibitem{leble}                  
Leble S.B., Vereshchagin D.A., Shchekin A.K. The kinetic
description of wave disturbances in the stratified gas. In "
Methods of hydrophysical researches", (1990.) P.215-233.
\bibitem{veresh}                 
Vereshchagin D.A., Shchekin A.K., Leble S.B. Boundary regime
propagation in a stratified gas with arbitrary Knudsen number.
Zhurnal Prikl.Mech. and Tehn.Fiz., $N^{\underline o} \ 5$.
P.70-79. (in Russian).
\bibitem{shchekin}               
Shchekin A.K., Leble S.B., Vereshchagin D.A. Introduction in
physical kinetic of rarefied gas. Kaliningrad. (1990). 80.p.(in
Russian).
\bibitem{lees}                   
Lees L. Kinetic theory description of rarified gas flow.
J.Soc.Industr. and Appl.Math.,13.$N^{\underline o} \ 1$. (1965).
P.278-311.
\bibitem{liu}                    
Liu Chung Yen., Lees L. in''Rarefied gas dynamics'' (ed.by
L.Talbot). Academic Press. (1961). P.391-428.
\bibitem{schidl}                 
Shidlovskij I.P. The introduction in rarefied gas dynamics.
Moscow, Nauka. (1965).220.p.(in Russian).
\bibitem{kostom}                 
Kostomarov J.A. Ing.Journ.,3. $N 3$. (1963). (in Russian).
\bibitem{mott}                   
Mott-Smith H.M. The solution of the Boltzmann equation for a shock
wave. Phys.Rev.,82. (1951).P.885-892.
\bibitem{nambu}                  
Nanbu K., Watanabe Y. Analysis of the internal structure of shock
waves by means of the exact direct-simulation method.
Rep.Inst.High Speed.Mech., 48. $N^{\underline o} \
366$.(1984).P.1-75.
\bibitem{sampson}                
Sampson R.E., Springer G.S. Condensation on and evaporation from
droplets by a moment method. J.Fluids.Mech. 36. part.3.
(1969).P.577-584.
\bibitem{ivchenko}               
Ivchenko I. J.Coll and Interf.Science. 120. $N^{\underline o} \
1$. (1987). P.1-7.
\bibitem{leble2}                 
Leble S.B., Vereshchagin D.A. Kinetic description of sound
propagation in exponentially stratified media. Advances in
Nonlinear Acoustic (ed.H.Hobaek).Singapore. World Scientific.
(1993). P.219-224.
\bibitem{veresc2}                
Vereshchagin D.A., Leble S.B. Proceedings of International
Symposium on Nonlinear Theory and its Applications "NOLTA
'93".(Hawaii,1993). 3. (1993). P.1097-1100.
\bibitem{veresc3}
Vereshchagin D.A.,  Leble S.B. Piecewise continuous partition
function and acoustics in stratified gas. {\it Nonlinear Acoustics
in Perspective, ed. R.Wei,} (1996),p.142-146.
\bibitem{veresc4}
Leble S.B, F.L. Roman, D.A. Vereshchagin and J.A. White. Molecular
Dynamics and Momenta BGK Equations for Rarefied Gas in Gravity
field.  in {\it Proceedings of 8th Joint EPS-APS International
Conference Physics Computing CYFRONET-KRAKOW, Ed. P.Borcherds,
M.Bubak, A.Maksymowicz} (1996), p.218-221.
\bibitem{kerzon}
Kerson Huang. Statistical Mechanics ,1963.\\
\end{thebibliography}
\end {document}